\documentclass[fleqn,usenatbib]{mnras}

\usepackage{newtxtext,newtxmath}

\usepackage[T1]{fontenc}

\DeclareRobustCommand{\VAN}[3]{#2}
\let\VANthebibliography\thebibliography
\def\thebibliography{\DeclareRobustCommand{\VAN}[3]{##3}\VANthebibliography}

\usepackage{graphicx}	
\usepackage{amsmath}	
\usepackage{color}
\newcommand{\tp}[1]{\bf{#1}}

\title[The precursor structure ]{The precursor structure in relativistic shocks}

\author[B. Skuratovsky et al.]{
Barel Skuratovsky,$^{1}$\thanks{E-mail:barel.skuratovsky@gmail.com}
Yuri Lyubarsky,$^{2}$
and Tsvi Piran$^{1}$
\\
$^{1}$Racah Institute of Physics,
The Hebrew University of Jerusalem, Jerusalem 91904, Israel\\
$^{2}$Physics Department, Ben-Gurion University, Be’er-Sheva 84105, Israel
}

\date{Accepted XXX. Received YYY; in original form ZZZ}

\pubyear{2023}

\begin{document}
\label{firstpage}
\pagerange{\pageref{firstpage}--\pageref{lastpage}}
\maketitle

\begin{abstract}
We present a common unifying macroscopic framework  for precursors 
in relativistic shock waves. These precursors transfer energy and momentum from the hot downstream to the cold upstream, modifying the shock structure. It was shown that in a steady state, there is a maximal fraction of the downstream energy flux that the precursor can carry.  We show here that at this critical value, the shock disappears, and the flow passes through a sonic point. This behavior resembles the classical Newtonian Rayleigh flow problem. At the critical value, the transition is unstable as perturbations in the upstream accumulate at the sonic point. Thus, if such a point is reached, the shock structure is drastically modified, and the flow becomes turbulent. 
\end{abstract}




\section{Introduction} \label{sec:intro} 
Relativistic shock waves are ubiquitous in numerous objects, ranging from AGNs to Gamma-Ray Bursts.  A common image of a shock is of an abrupt discontinuity with no information passing from the downstream to the upstream. However, in many cases, this picture is oversimplified. A  precursor can propagate ahead of the shock into the upstream.  Precursors that can overtake the shock include fast particles (sometimes called cosmic rays), radiation (as in the case of radiation-mediated shocks), or supersonic plasma waves. If sufficiently strong, they can significantly modify the upstream and, consequently, the overall shock structure. 

The role of fast particles as precursors of  Newtonian shocks has been extensively studied since the classical work of \cite{Bell2004}. It was later extended to relativistic shocks  \cite[e.g.][]{Couch2008,Nakar2011,lemoine_physics_2019}. 
In radiation-mediated shocks,  photons transfer energy from the downstream to the upstream. {There}, the shock structure is completely dominated by this process, and the  width of the region is the mean free path of the photons. Detailed studies of such shocks have been carried out both numerically and analytically  \citep{Budnik2010,Nakar2012,Granot2018}.
A different precursor was considered within the ``pair-balance" model  \citep{derishev_particle_2016}. In this model, high-energy photons generated in the downstream  annihilate with low-energy ones in the upstream near the shock, transferring energy and momentum to the upstream and modifying its structure. \cite{Garasev2016} have shown using numerical simulations that this may lead to a Weibel instability and generation of magnetic fields upstream. 
Another modification of the upstream structure can arise from 
low-frequency electromagnetic waves emitted by magnetized relativistic shocks \citep{Lyubarsky2018}. In particular, such an interaction in electron-ion flows leads to electron heating  and even to the non-thermal particle acceleration \citep{Lyubarsky2006,Hoshino2008,Sironi2011,Iwamoto2022}. 

All  these phenomena   incorporate energy and momentum transfer  from the downstream to the upstream. 
Often, these fluxes are negligible, but at times, as in radiation-mediated shocks and in the ``pair balance" model, they are important and even dominant. 
We discuss here a general framework for the macroscopic modification of a shock wave profile due to energy and momentum transfer from the shock's downstream to the upstream.   In \S \ref{steady state} we describe,  following \cite{derishev_particle_2016}, the steady state problem for a relativistic shock.   
There is a maximal fraction of the total energy that can be transferred from the downstream to the upstream in a steady-state solution. 
This result  resembles the Newtonian Rayleigh flow problem, { Concerned with a flow modified by heat deposition (or removal), in which there is also a maximal amount of heat that can be added (or removed) from the flow. At the maximum, the shock wave disappears completely, and the solution  passes through a sonic point. } In 
\S \ref{sec:stability_rel} we show that as the flow velocity approaches the speed of sound (from above), instability develops. 
Possible implications of our results to different astrophysical models and, in particular, of the turbulent flow near the critical point for magnetic field build-up and particle acceleration are discussed at \ref{sec:discussion}.

\section{Steady state solution}
\label{steady state}

\begin{figure}
    \centering
    \includegraphics[scale=0.45]{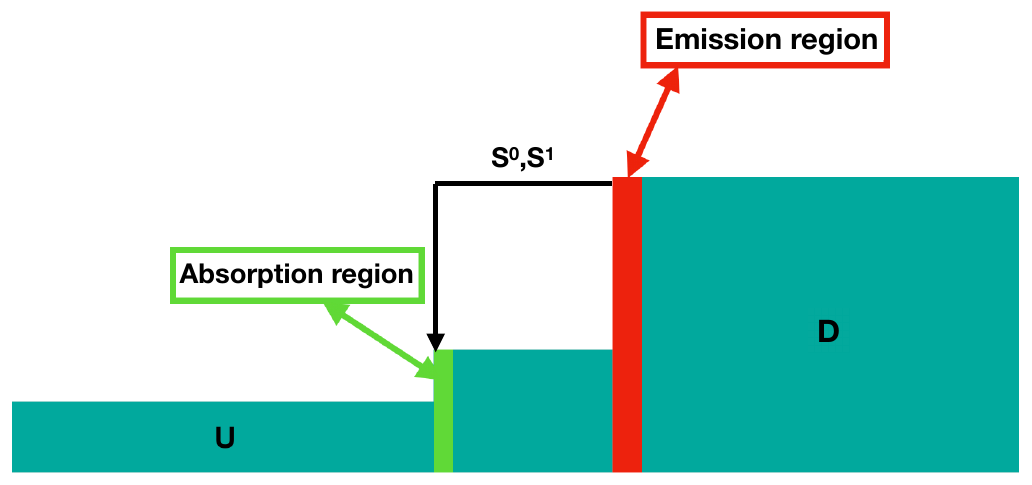}
    \caption{A schematic description of the energy $S^0$ and momentum $S^1$ fluxes from the downstream (where the extraction region is marked in red) to a point in the upstream (marked in green). Generally, the energy and momentum deposition are not limited to a single location, and the fluxes are a function of the distance from the shock.  }
    \label{fig:Schematic}
\end{figure}

We consider the energy and momentum transfer from the shock downstream to the upstream. The process is studied in the shock frame. The overall structure of the flow is the following (see Fig. \ref{fig:Schematic}). The shock is placed at $x^1=0$, and the flow is directed towards { positive}  $x$. At the far upstream (formally at $x^1\to  \tp{-}\infty$) the flow is cold. The relativistically hot plasma downstream emits a fraction of its energy upstream in the form of radiation or relativistic particles. This energy is absorbed in an extended region upstream of the shock so that the flow is heated and decelerated. As a result, the shock discontinuity decreases and may even disappear completely. Global energy conservation implies that far downstream (formally at $x^1\to \infty)$, the flow parameters are related to the far upstream parameters by the standard Taub adiabat. We explore here how the abosrption of energy and momentum affects the upstream flow and of the shock wave.

The  perfect fluid is described by an  energy-momentum tensor
\begin{equation}
T^{\mu\nu}=wu^{\mu} u^{\nu}+g^{\mu\nu}p \ , 
\end{equation}
where $p$ is the pressure, $w\equiv e+p$ is the enthalpy,  $e$ is the energy density, $u^{\nu}$ is the four-velocity and  $g^{\mu\nu} \equiv \eta^{\mu\nu}$ is the flat space metric. 
The motion is one-dimensional and stationary in the rest frame of the shock. In this case $u^{\mu}(x)= (\Gamma(x), \Gamma (x) \beta (x), 0,0)$, where we denote by $x \equiv x^1$ the spatial coordinate along which the shock moves. 
The transfer of energy and momentum from the downstream to the upstream  is described by the vector function $S^{\mu} (x) \equiv  (S^{\rm 0}(x) ,S^{\rm 1}(x),0,0)$, the components of which are defined as the energy and momentum fluxes emitted from the downstream that are present at the upstream at distance $|x|$ ahead of the shock. { Within the downstream  $S^{\mu} (x)$  describes the build up of these fluxes. }
The equations of motion are written as: 
\begin{equation}
\frac{\partial T^{\mu\nu}}{\partial x^{\nu}}=\frac{d{S^{\mu}}}{{dx}} \ .
\label{eq.EOM} 
\end{equation}
These equations  are supplemented by the continuity of the baryon number density:
\begin{equation}
\frac{\partial (n u^{\mu})}{\partial x^{\mu}}=0 \ .
\label{eq:n} 
\end{equation}

The  { returning} energy flux,  $S^0$, is conveniently normalized by the total upstream energy density flux:  
\begin{equation} 
S^{\rm 0} (x)\equiv Q(x) T^{01}_{\rm u}  \ , 
\label{eq:Se}
\end{equation}
where the subscript $u$  denotes the far upstream  region, where no energy has been absorbed yet. Recall that the energy fluxes far upstream and far downstream are equal.
The function $Q(x)$ defines the fraction of the downstream energy flux that reaches a point distanced $x$ from the shock. $Q(-\infty)=0$ (as we assume that all the escaping flux is absorbed)\footnote{{ By choosing $ Q(-\infty) \ne 0$ our results can be easily generalized to the case that some of the flux escapes.}}. 
{ $Q_0 \equiv Q(0)$} is the total fraction of energy that is extracted from the downstream. 
In the upstream $Q(x) $ { increases} 
{ (i.e. it increases with decreasing x as we approach the shock)}
as less and less flux reaches further away from the shock.
{ If considered as a function of $\Gamma\beta$, $Q(\Gamma\beta$ decreases in the upstream.} 
$Q(x)$ { decreases (with x)} in the downstream as more and more flux is emitted, and it reaches a maximum, $Q_0$ at the shock. 

{ We define $\eta(x)$ as the local ratio of the returning}  momentum flux, $S^{\rm 1}(x)$,  to the  { returning} energy flux:
\begin{equation} 
S^{\rm 1} (x)\equiv - {\eta (x)}{\beta_{u}} S^{\rm 0}(x)  \ .
\label{eq:Sp}
\end{equation}
{ We define $\eta_0\equiv \eta(0)$ as the ratio of the fluxes at $x=0$.}

We describe the system as composed of two fluids: Baryon fluid that dominates the mass density and remains cold in the upstream flow, and radiation and lepton fluid that could dominate the energy density and pressure. In addition to the original electrons in the system, pairs can be produced in the absorption region. The leptons are relativistic, and their equation of state that is combined with the radiation equation of state satisfies:  
\begin{equation}
p=\frac{e_l}{3} = \frac{w_l}{4} \  , 
\label{eq: e.o.s}
\end{equation} 
where we ignore the mass of the leptons.
The far upstream flow is cold, such that:
\begin{equation}
w_{\rm u} = n_{\rm u} \ . 
\end{equation}
We use units in which the proton's rest mass and the speed of light are unity,  $m_p=c=1$. In the absorption region,  we have  a combination of baryons (whose thermal energy and pressure are neglected), leptons (whose rest mass is neglected), and radiation.
The upstream enthalpy is written as
\begin{equation}
    w=n+\frac 43e_l \ .
\label{eq: e.o.s1}\end{equation}

Under these assumptions, the conservation laws are: 
\begin{equation}
    n\beta\Gamma=n_u\beta_u\Gamma_u \ ,
\label{eq:cont}
\end{equation} 
\begin{equation}
    w\beta{\Gamma^2}
    =n_{\rm u}\beta_{\rm u}{\Gamma^2_{\rm u}} \left[1+Q(x)\right] \ , 
    \label{eq:energy_equation}
\end{equation}
\begin{equation}
p+w (\beta \Gamma)^2=n_{\rm u} (\beta_{\rm u} \Gamma_{\rm u})^2[1 -{ \eta}Q(x)] \ .
    \label{eq:rel_momentum}
\end{equation}
{ Note that the sign of the returning momentum flux is opposite to the one of the energy flux.}

Eliminating  $n$, $p$ and $w$, we obtain:
\begin{equation} 
    Q(x)=\frac {({{1}+ 4 \Gamma_{\rm u} \beta_{\rm u } \Gamma \beta}) \Gamma - {({1}+4{\Gamma^2}\beta^2})  {\Gamma_{\rm u}}}
     {4 \eta(x) {{  \Gamma_{\rm u}  \Gamma^2 \beta {\beta_{u}} }} + ({{1}+4{\beta^2}{\Gamma^2})}{\Gamma_{\rm u}}  }\ .
   \label{eq:expression_a}
\end{equation}
This equation relates the four-velocity at any point to the ``returning"
flux at this point (measured in units of the upstream energy flux), $Q(x)$, { and to the momentum flux (via $\eta(x)$}).  Once $  \Gamma$ is known, we determine $n$ from the continuity equation and  $w$ from Eq. \eqref{eq:energy_equation}. 
\begin{figure}
    \centering
    \includegraphics[scale=0.5]{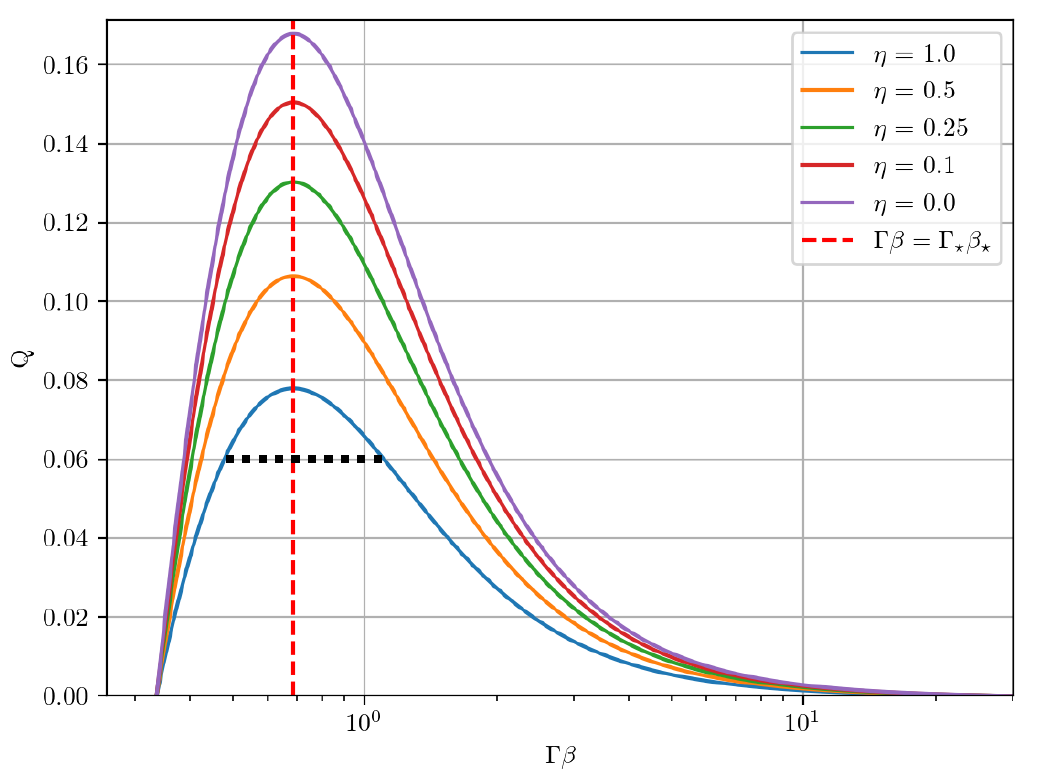}
    \caption{$Q$ vs. $\beta\Gamma $ for   different values of $\eta$ (for $\Gamma_{\rm u} \beta_{\rm u} =30$). The red line marks the speed of sound for $\eta=0$,  defined at $dQ=0$. $\beta_{\star}={1}/{\sqrt{3}}$ is a good approximate for $\beta$ for which $Q$ is maximal. { The region to the right of the maximum corresponds to the upstream. In this region as we go further into the upstream ($\Gamma \beta$ increases) the flux $Q$ decreases. The region to the left of the maximum corresponds to the downstream. In this region the flux is building up and as $\Gamma \beta$ increases  the flux $Q$ increases. The dotted line depicts the shock in a solution with $Q_0=0.06$ and $\eta_0=1.$. In this case h $\Gamma\beta$ jumps at the shock from $\approx 0.45$ to $\approx 1.1$.}}
    \label{fig: Q_chi_upstream_30}
\end{figure}

Fig. \ref{fig: Q_chi_upstream_30}, depicts  $Q(\Gamma\beta)$  for $\Gamma_{\rm u } \beta_{\rm u}=30 $ for several values of { constant}  $\eta$  \citep[See also Fig. 2 of][]{derishev_particle_2016}. First, we notice that momentum transfer changes quantitatively but not qualitatively the results.
Notably, the maximal value of $Q$, $Q_{\rm max }$ decreases monotonically as $\eta $ increases.  For $\Gamma_{u}\beta_{u}=30$, the maximal value of energy transferred 
is only $\approx0.17$. As we show below when the energy transfer is maximal, the flow velocity at the location of the shock decreases to the local speed of sound. In this case, the shock disappears, and the flow is continuously decelerated. The part of the curve to the right of the maximum describes the  deceleration of the upstream from the initial velocity. The part to the left of the maximum describes the  deceleration in the downstream until the Taub adiabat determines the velocity. 

{ The trajectories shown in Fig. \ref{fig: Q_chi_upstream_30} are for constant values of $\eta$. If $\eta(x)$ varies, namely if the fraction of momentum absorbed is not proportional to the fraction of energy absorbed, the solution will move from one constant $\eta$ trajectory to another. As $Q(x)$ cannot increase with increasing $\Gamma \beta$, that is as we move towards the far upstream, there is a limit on how $\eta (x)$ can vary. This limit  can be derived by demanding, using  Eq. \ref{eq:expression_a}, that $dQ/dx \ge 0$.   
}

For a given $Q_0 < Q_{\rm max}$ { and a given $\eta_0$,  the $\Gamma \beta  $ values corresponding to the intersection of $Q_0$ with the $Q(\Gamma \beta)$ curve are the velocity jump at the shock. The  $Q(\Gamma \beta)$ curve  to the right of intersection of the line $Q=Q_0$ with the $Q(\Gamma\beta)$ curve describes the upstream flow, in which energy is absorbed and $Q$ decreases.  The  curve to the left of the intersection of $Q=Q_0$ with $Q(\Gamma\beta)$ curve  corresponds to the downstream regime in which the energy extraction is building up, from zero at the distant downstream up to $Q_0$ at the shock discontinuity. }

\begin{figure}
    \centering
    \includegraphics[scale=0.5]{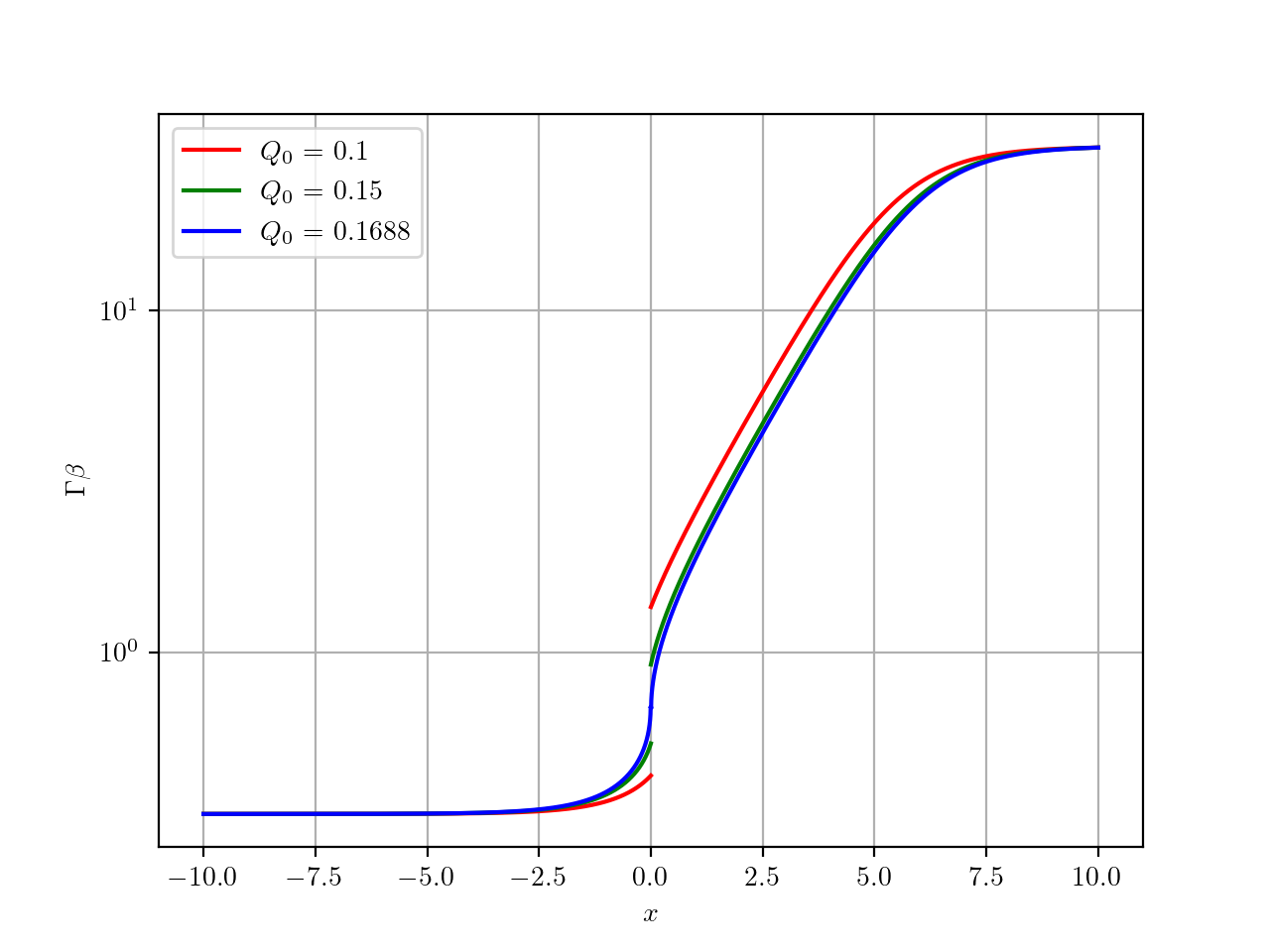}
    \caption{$\Gamma(x)\beta(x)$ for an absorption law of \eqref{eq:abs_prof}, $\Gamma_{u}\beta{u}=30$ and  $\eta=0$. { the jump at the shock depends on the value of $Q_0$.}} 
    \label{fig: chi_x_upstream}
\end{figure}

As an example of a possible structure of the upstream (and the downstream),  Fig. \ref{fig: chi_x_upstream} depicts the modified shock profile  as a function of $x$, given an absorption (and emission) profiles  \begin{equation}
Q(x) = Q_0 \cdot      \begin{cases} 
     e^{x}\hspace{0.7cm}x<0 \ , \\
     e^{-x}\hspace{0.5cm}x>0 \ .
\end{cases}
    \label{eq:abs_prof}
    \end{equation}
{In the downstream region $(x>0)$,  $Q(x)$ describes the build-up of the flux due to emission in this region. In the upstream region it describes the decreasing absorbed flux.}
Note that this absorption law (Eq. \ref{eq:abs_prof}) is schematic, and generally relativistic effects make the absorption law much more complicated \cite[see e.g.][]{Granot2018}. In spite of its simplicity, the resulting profile resembles those found numerically by \cite{lemoine_physics_2019} in PIC simulations in which high energy particles accelerated in the downstream deposit their energy in the upstream. { These authors \cite{lemoine_physics_2019} also solved analytically a similar modified shock model. }

\begin{figure}
    \centering
    \includegraphics[scale=0.5]{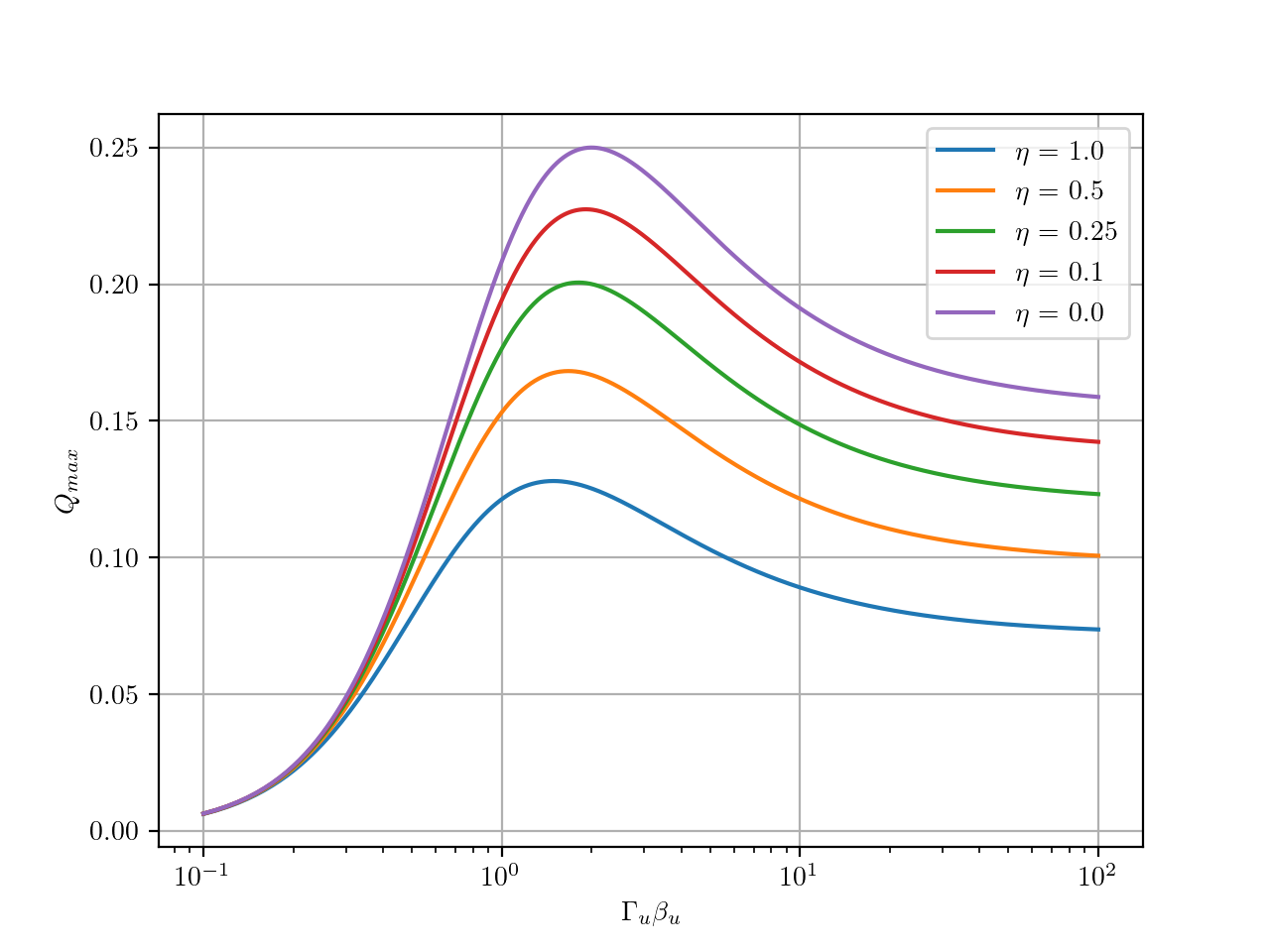}
    \caption{$Q_{max}$ as a function of $\Gamma_{u}\beta_{u}$. The maximal value of $Q_{max}$,  ${1}/{4}$, is achieved when $\Gamma_{u}\beta_{u}={2}$ and $\eta=0$.}  
    \label{fig: Q_critical_chi_u}
\end{figure}

The value of  $Q_{\rm max }$ as a function of $\eta_0$ and the upstream conditions is  given by: 
\begin{equation}
    Q_{\rm max}\left(\eta_0, \beta_{u}\right) = \frac{Q_{\rm max}\left(\eta_0, \beta_{u}\right)}{1+{4\eta_0{\beta_{\star}}\beta_{u}}/({1+3\beta^2_{\star}})}
\end{equation}
{ Here $\beta_{\star}$ is the velocity at the point of $Q=Q_{\rm max }$.}
Fig. \ref{fig: Q_critical_chi_u} depicts $Q_{\rm max }$ as a function of $\Gamma_u \beta_u$ for several values of $\eta_0$.  One can see that it is maximal at $ \Gamma_{u}\beta_{u} \approx 2$. If $\Gamma_u\gg 1$, the downstream flow becomes relativistically hot, and the speed of sound approaches $\beta_s = 1/\sqrt 3$. As $\Gamma_{u}\to\infty$, the maximum of Eq.\eqref{eq:expression_a} is achieved at    $ \beta={1}/{\sqrt{3}}$. At this limit, 
 \begin{equation}  \lim_{\Gamma_{u}\rightarrow{\infty}}{Q_{\rm max}}=\frac{2-\sqrt{3}}{\sqrt{3}+2\eta_0}
\label{Qmax} \end{equation}

\begin{figure}
    \centering
    \includegraphics[scale=0.5]{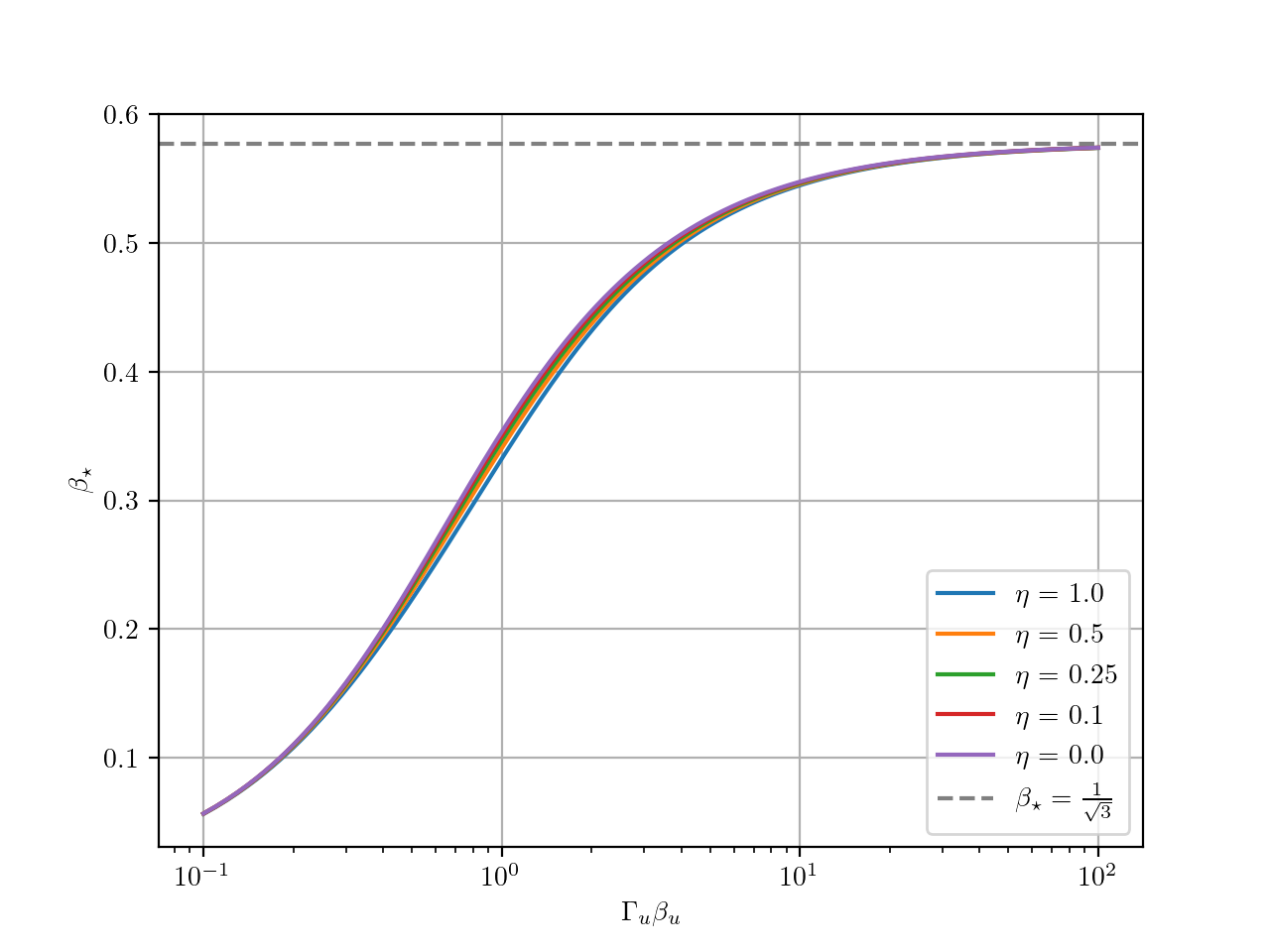}
    \caption{
    The velocity at the sonic point as a function of $\Gamma_{u}\beta_{u}$ for different values of $\eta$.} 
    \label{fig: speed_of_sound_no_title}
\end{figure}

{ Let us show that the flow velocity at the point $Q=Q_{\rm max}$
is equal to the local sound speed. Differentiating Eqs. \eqref{eq:cont}-\eqref{eq:expression_a} and taking into account that the equation of state is given by Eq. \eqref{eq: e.o.s} and that $dQ=0$ at the point of interest, we get
\begin{gather}
\beta_\star dn+n_\star\Gamma_\star^2d\beta=0;\\
\beta_\star dw+w_\star(1+\beta_\star^2)\Gamma_\star^2d\beta=0;\\
-\frac 14dn+\left(\frac 14+\beta^2_\star\Gamma_\star^2\right) dw+2w_\star\beta_\star^2\Gamma_\star^4d\beta=0.
\end{gather}
Here, the index star is referenced to the quantities at this point.
The obtained homogeneous set of equations has a nontrivial solution if the determinant vanishes, which implies
\begin{equation}
    3\beta_{\star}^2=1-\frac{n_\star}{w_\star}. 
\end{equation}
On the other hand, the speed of sound in the medium with the equation of state \eqref{eq: e.o.s} is found as
\begin{equation}
\beta_s^2=\left(\frac{\partial p}{\partial e}\right)_S=\frac 13\left(1-\frac nw\right).
\end{equation}
One sees that the velocity of the flow achieves the local speed of sound at the point where the absorbed energy is maximum. 

Applying the condition $dQ=0$ to Eq. \eqref{eq:expression_a}, one finds an equation for $\beta_\star$:}
\begin{equation}            
    \Gamma^3_{\star}(\beta^3_{\star}+\eta\beta_{u})+\Gamma_{u}\beta_{u}(2\Gamma^2_{\star}\beta^2_{\star}-1)(1+\eta)+\frac{7}{4}\Gamma_{\star}\beta_{\star}=0 \
    \label{eq:speed_of_sound2}
\end{equation}
For the highly relativistic upstream motion, $\Gamma_u \rightarrow \infty$ this equation yields, as expected, $\beta_\star \rightarrow \frac 1{\sqrt{3}}$. Fig.\ \ref{fig: speed_of_sound_no_title} depicts the dependence of the flow velocity at the sonic point on the upstream velocity. 

There is no steady state solution with $Q>Q_{\rm max}$. We expect that if this value is reached than either the downstream readjust itself and the emissivity is reduced or a non steady state turbulent solution develops.  

It is interesting to note that this  relativistic solution  is equivalent to the classical Rayleigh flow \citep[e.g.][]{emanuel_gasdynamics_1986} that describes the evolution of the flow  when heat is added or removed from a compressible ideal Newtonian fluid in a constant area duct. The flow chokes thermally if too much energy is added. This happens at a sonic point that corresponds to the critical value of heat deposition. In our setting the modified upstream is equivalent to a supersonic Rayleigh flow with heat deposition.

\section{Instability of the continuous solution}
\label{sec:stability_rel}
In the previous section, we found a steady-state solution  only if the fraction of the energy transferred from the downstream to the upstream flow does not exceed the value given by Eq.\ (\ref{Qmax}). The upstream velocity, and therefore the shock discontinuity, decreases when the absorbed energy increases. At $Q=Q_{\rm max}$, the flow velocity reaches the sound speed at $x=0$ so that the transition to the subsonic downstream flow becomes continuous. The absence of steady-state solutions at $Q>Q_{\rm max}$ implies that at this condition, the flow becomes unsteady. This conjecture is supported by the fact that the continuous solution at $Q=Q_{\rm max}$ is in fact unstable: the amplitude of small sound perturbations  diverges when the flow approaches the sound point. The reason is that the sound perturbation is dragged forward by the flow. Therefore the velocity of perturbations, that are directed backward in the flow frame, drops to zero at the sound point, which implies that the amplitude diverges. In this section, we present the formal derivation of this instability. 

For simplicity, we assume that $\Gamma_u\gg 1$. In this case, most of the energy is converted into the thermal energy of leptons in the vicinity of the sound point where the flow is only mildly relativistic. Therefore one can neglect the contribution of cold protons into the equation of state and write $w=4p$. In this case, the flow is described only by the energy and momentum equations \eqref{eq.EOM}, whereas using the continuity equation is not necessary. 
We project the equations of motion \eqref{eq.EOM} on the 4-velocity and on the perpendicular direction. For this purpose, make a dot product of the equations with $u_\nu$ and with the projection operator $g_{\mu\nu}-u_\mu u_\nu$. This yields
\begin{gather}
\frac{\partial wu^\mu}{\partial x^\mu}-u^\mu\frac{\partial p}{\partial x^\mu}=\frac{d{S^{\mu}}}{dx^{1}}u_\mu\\
wu^\mu\frac{\partial u_\nu}{\partial x^\mu}-\frac{\partial p}{\partial x^\nu}+u_\nu u^\mu\frac{\partial p}{\partial x^\mu}=\frac{d{S_{\nu}}}{dx^{1}}-u_\nu u_\mu{\frac{d{S^{\mu}}}{dx^{1}}}
\end{gather}
We assume for simplicity that the momentum source is much smaller than the energy source so that $S^{1}=0$. Then we get for a one-dimensional flow
\begin{gather}
4\left(\frac{\partial p\Gamma}{\partial t}+\frac{\partial p\Gamma\beta}{\partial x}\right)-\Gamma\frac{\partial p}{\partial t}-\Gamma\beta\frac{\partial p}{\partial x}=\Gamma \frac{\partial S^{0}}{\partial x}\\
4p\left(\frac{\partial \Gamma\beta}{\partial t}+\beta\frac{\partial \Gamma\beta}{\partial x}\right)+\Gamma\beta\frac{\partial p}{\partial t}+\Gamma\frac{\partial p}{\partial x}=-\Gamma\beta \frac{\partial S^{0}}{\partial x} \ . 
\end{gather}
In the steady state, these equations are reduced to
\begin{gather}
4\bar p\bar \Gamma^2\bar \beta'+3\bar\beta \bar p'=S'^{0};\label{steady1}\\
4\bar p\bar\Gamma\bar\beta^2\bar\beta'+\bar p'=-\bar\beta S'^{0}\label{steady2} \ ,
\end{gather}
where prime denotes a $x$ derivative, and we used bar to denote parameters in the steady state.
Eliminating $\bar p'$ from these equations yields
\begin{equation}
4\bar p(1-3\bar\beta^2)\bar\Gamma^2\bar\beta'=(1+3\bar\beta^2) S'^{0},
\label{source}
\end{equation}
which means that the flow passes the sonic point, $\beta=1/\sqrt{3}$, only if $S'_{e}$ vanishes at this point, consistent with our findings about the critical point. 

To study the stability of the steady-state flow, we consider a small perturbation to the steady state solution, $\beta={\bar\beta}+\delta{\beta}$ and $p={\bar p}+\delta{p}$, where $\delta\beta\ll {{\bar\beta}}$; $\delta{p}\ll{{\bar p}}$. Linearizing the equations, we get:
\begin{gather}
3\left(\frac{\partial \delta{p}}{\partial t}+{\bar\beta}\frac{\partial \delta{p}}{\partial x}\right)+4{\bar p}{\bar\Gamma}^2\left({\bar\beta}\frac{\partial \delta\beta}{\partial t}+\frac{\partial \delta\beta}{\partial x}\right)+\nonumber \\ 4{\bar\beta'}{\bar\Gamma^2}\delta{p} 
+(3{\bar p'}+8{\bar p}{\bar\beta}{\bar\beta'}{\bar\Gamma^4})\delta\beta=0 \ ,
\label{1}\\
{\bar\beta}\frac{\partial \delta{p}}{\partial t}+\frac{\partial \delta{p}}{\partial x}+4{\bar p}{\bar\Gamma}^2\left(\frac{\partial \delta\beta}{\partial t}+{\bar\beta}\frac{\partial \delta\beta}{\partial x}\right)+\nonumber \\ 4{\bar\beta'}\left[{\bar\beta}{\bar\Gamma^2}\delta{p}+{\bar p}{{\bar\Gamma}^4}(1+{\bar\beta}^2)\delta{\beta}\right]=-\delta{\beta}S'^{0} \ .
\label{2}
\end{gather}
Assuming high frequency oscillations, such that $c/\omega$ is much smaller than the characteristic scale of the flow, {we employ the WKB approximation. For the non-relativistic Rayleigh flow, the procedure has been developed by \citet{umurhan_wkb_1999}.}

Expanding the solution in powers of $1/\omega$ we obtain:
\begin{gather}
\begin{pmatrix}
\delta{\beta} \\ \delta{p}
\end{pmatrix}=
\begin{pmatrix}
\beta_0+{\beta_1}/{\omega}+\dots \\ p_0+{p_1}/{\omega}+\dots
\end{pmatrix}\exp\left[i\omega\left(-t+\int\frac{dx}u\right)
\right],
\label{ansatz}\end{gather}
where $u$ is the wave phase velocity. Substituting this ansatz into Eqs. (\ref{1}) and (\ref{2}) and collecting the terms of the order of $\omega$ one gets
\begin{gather}
3\left(-1+\frac{{\bar\beta}}{u}\right)p_0+4{\bar p}{\bar\Gamma}^2\left(-{\bar\beta}+\frac{1}{u}\right)\beta_0=0 \ , \label{zeroth_order1}\\
\left(-{\bar\beta}+\frac{1}{u}\right)p_0+4{\bar p}{\bar\Gamma^2}\left(-1+\frac{{\bar\beta}}{u}\right)\beta_0=0 \ .\label{zeroth_order2}
\end{gather}
This set of equations has a nontrivial solution if
\begin{equation}
u=\frac{{\bar\beta}\pm 1/\sqrt{3}}{1\pm {\bar\beta}/\sqrt{3}} \ .
\label{velocity}\end{equation}
We see that the wave velocity is a relativistic superposition of the flow velocity, ${\bar\beta}$, and the sound velocity, $\pm 1/\sqrt{3}$. Substituting $u$ back to Eq.\ \eqref{zeroth_order2} we find a relation between $p_0$ and $\beta_0$:
\begin{equation}
p_0=\pm\frac 4{\sqrt{3}}{\bar p}\bar\Gamma^2{\beta_0} \ .
\label{d0}\end{equation}
In the next approximation, one collects terms of the order of $\omega^0$. This yields a set of equations of the form
\begin{eqnarray}
&&i\hat{\mathbf M} \begin{pmatrix} \beta_1 \\ p_1\end{pmatrix} ={\mathbf\Phi} = -
\\
&&\begin{pmatrix}
3{\bar\beta} p'_0+4{\bar p}\bar\Gamma^2 \beta'_0+4\bar\beta'\bar\Gamma^2{p_0}+(3\bar{p}'+8{\bar p}{\bar\beta}\bar\beta'\bar\Gamma^4)\beta_0 \\
p'_0+4{\bar p}{\bar\beta}\bar\Gamma^2\beta'_0+4\bar\beta'\left[{\bar\beta}\bar\Gamma^2p_0+{\bar p}(1+\bar\beta^2)\bar\Gamma^4{\beta_0}\right]+\beta_{0}S'^{0} 
\end{pmatrix} \ ,
\nonumber 
\label{eq:eqs1}
\end{eqnarray}
where the matrix $\hat{{\mathbf M}}$ is the same as in  Eqs. (\ref{zeroth_order1},\ref{zeroth_order2}). 
Substituting $u$ from Eq. (\ref{velocity}), we find
\begin{equation}
\hat{{\mathbf M}}=\frac 1{\bar{\Gamma^2}({\bar\beta}\pm 1/\sqrt{3})}\begin{pmatrix} \mp \sqrt{3} & 4{\bar p}\bar{\Gamma^2}\\ 
 1 & \mp \frac 4{\sqrt{3}}{\bar p}\bar{\Gamma^2}\end{pmatrix}.
\end{equation}
Substituting  $\bar p'$ from \eqref{steady2},  $p_0$ from \eqref{d0} and $S'_{e}$ from \eqref{source} into $\mathbf\Phi$ yields:
\begin{gather}
    {\mathbf\Phi}=- 4 {\bar p}\bar\Gamma^2 \begin{pmatrix}
(1\pm \sqrt{3}{\bar\beta}) \beta'_0 + \bar\beta'
\left\{\frac{2\left({\bar\beta}\pm\frac{2}{\sqrt{3}}\right)}{1-{\bar\beta}^{2}} - \frac{2\left({{\bar\beta}} \pm{\sqrt{3}\bar\beta^2}\right)}{{\bar\beta}^{2}+\frac{1}{3}}
 \right\}\beta_0\\
({\bar\beta}\pm \sqrt{3}) \beta'_0+\bar\beta'\left\{\frac{{\bar\beta}^{2}\pm\frac{4}{\sqrt{3}}{\bar\beta}+1}{1-{\bar\beta}^{2}}-\frac{{\bar\beta}^{2}\pm\frac{2}{\sqrt{3}}{\bar\beta}-\frac{1}{3}}{{\bar\beta}^{2}+\frac{1}{3}}\right\}\beta_0
\end{pmatrix}
\end{gather}
The matrix $\hat{{\mathbf M}}$ is degenerate;  the condition that  Eqs.  \ref{eq:eqs1} has a solution is
\begin{gather}
{\rm Det}\begin{pmatrix}
\mp \sqrt{3} & -(1\pm \sqrt{3}{\bar\beta})\beta'_0+
\bar\beta'\beta_{0}\left\{\frac{2({\bar\beta}\pm{2}/{\sqrt{3}})}{1-{\bar\beta}^{2}} -
\frac{2({\bar\beta}\pm \sqrt{3}{\bar\beta^2})}
{{\bar\beta}^{2}+{1}/{3}}\right\} \\
1 & -({\bar\beta}\pm \sqrt{3})\beta'_0-\bar\beta'\beta_{0}\left\{
\frac{2({\bar\beta}\pm{2}/{\sqrt{3}})}{1-{\bar\beta}^{2}} - \frac{ 2({\bar\beta} \pm\sqrt{3}{\bar\beta^2})}{{\bar\beta}^{2}+{1}/{3}}\right\}
\end{pmatrix}=0
\label{deter}
\end{gather}
Choosing the lower sign of $\pm$ in \eqref{deter}, which describes the  wave propagating backwards in the comoving frame, leads to 
\begin{equation}
    \beta'_{0}({\bar\beta}-\frac{1}{\sqrt{3}})=-{\bar\beta}'\beta_{0}\left\{\frac{\frac{{\bar\beta}^{2}}{2}-\sqrt{3}{\bar\beta}+\frac{7}{6}}{1-{\bar\beta}^{2}}-\frac{\frac{3}{2}{\bar\beta}^{2}-\frac{2}{\sqrt{3}}{\bar\beta}-\frac{1}{6}}{{\bar\beta}^{2}+\frac{1}{3}}\right\}
    \label{final_eq}
\end{equation}
Integrating \eqref{final_eq} yields 
\begin{equation}
    \beta_{0}=\frac{C}{\bar\Gamma}\cdot\frac{\bar\beta^2+ \frac{1}{3}}{\bar\beta-\frac{1}{\sqrt{3}}}\cdot{\left(\frac{1+\bar\beta}{1-\bar\beta}\right)}^{{\frac{1}{\sqrt{3}}}}, 
\end{equation}
where C is an integration constant. Since $p_{0}$ is proportional to $\beta_{0}$ by Eq. (\ref{d0}), we conclude that at the zeroth order, both $\delta\beta$ and $\delta{p}$ diverge as ${\bar\beta}$ approaches the speed of sound from above, meaning that transition through the critical point is unstable.
The result can be understood physically. Perturbations in the upstream propagate at the speed of sound both towards and away from the shock. As long as the shock speed is supersonic, perturbations from the upstream won't reach the shock. However, as we reach the critical energy transfer, the shock speed approaches the speed of sound. At this stage, perturbations that move towards the shock reach it and accumulate there, resulting in overall instability. 

\section{Discussion}
\label{sec:discussion}
We presented a common framework  unifying the description of macroscopic precursor phenomena in relativistic shock waves. Our general results for the perturbed upstream structure agree nicely with those obtained in numerical PIC simulationa \cite[e.g.][]{lemoine_physics_2019} in which high energy particles accelerated in the downstream deposit their energy in the upstream. In  a steady state, there is an upper limit to the fraction of energy transferred from the downstream to the upstream \citep{derishev_particle_2016}. This maximal energy  decreases when the   momentum  transferred increases.  It approaches a constant value, $(2-\sqrt{3})/(\sqrt{3}+2\eta)\approx 0.155/(1+1.15 \eta)$, in the extreme relativistic case, when the Lorentz factor of the cold upstream approaches infinity.

There are several implications to this result. The first corollary deals with radiation-mediated shocks, in which energy transport  from the downstream to the upstream is a dominant phenomenon. 
{Typically, relativistic radiation mediated shocks involve a collisionless subshock. Our analysis explains this phenomenon. The radiation flux from the downstream may not  be fine-tuned to the value required for a smooth transition. It is important to stress that for radiation-mediated shocks our solution is valid only for the upstream region. In this region any photon that interacts with the relativistic flow creates a pair that moves with the flow.  Indeed, in this region our solution is consistent with those of \cite{budnik_relativistic_2010} and \cite{Granot2018}. However, the downstream is mildly relativistic and there the photons' behavior is diffusive. For this reason Fig. \ref{fig: Q_chi_upstream_30} cannot be used to estimate  the subshock jump in these shocks. Moreover, the stability analysis discussed in \S \ref{sec:stability_rel} may not be applicable.   }

A second novel result is that once the energy transport to the precursor reaches a maximal value, the shock disappears, and the solution passes through a sonic point. However, this marginal solution is unstable. The physical reason for the instability is apparent. Perturbations in the upstream move at the speed of sound relative to the local flow. When the upstream velocity ahead of the shock approaches the speed of light, the "outgoing" perturbation directed towards the far upstream is carried towards the sonic point. As they cannot cross the sonic point, they accumulate there, resulting in instability.

This instability implies that if, for some reason, the energy flux from the hot downstream is too large, it will induce unstable turbulence in the upstream. The characteristic  scale  of this turbulence will be macroscopic, as it should correspond to  the scale in which the energy flux is deposited in the upstream. One can speculate that such large-scale turbulence could be the source of large-scale magnetic fields.  If correct, this will resolve one of the puzzles involving collisionless shocks in GRB afterglows - what is the origin of the large scale (much larger than the local plasma skin depth) magnetic fields that are implied from the afterglow observations \citep{Gruzinov1999}. 

\section*{Acknowledgements}

We thank Evgeny Derishev, Ehud Nakar and Re'em Sari for helpful discussions and an anonymous referee for helpful remarks. This work was supported by an Advanced ERC grants TReX and MultiJets and by ISF grants 2126/22 (TP) {and 2067/19 (YL).}

\section*{Data Availability}
No new data has been generated. 

\end{document}